\documentstyle[aps,prl,epsf]{revtex}
\begin{document}
\draft \twocolumn[\hsize\textwidth\columnwidth\hsize\csname
@twocolumnfalse\endcsname 
\title{{\bf Solitons in the one-dimensional
forest fire model}} 
\author{Per Bak$^{1,2}$, Kan Chen$^1$, and Maya
Paczuski$^{1,2}$} 
\address{Department of Computational Science,
National University of Singapore, Singapore 119260 \\ Department of
Mathematics, Huxley Building, Imperial College of Science, Technology,
and Medicine, London UK SW7 2BZ }

\date{\today }
\maketitle

\begin{abstract}

Fires in the one-dimensional Bak-Chen-Tang forest fire model propagate
as solitons, resembling shocks in Burgers turbulence. The branching of
solitons, creating new fires, is balanced by the pair-wise
annihilation of oppositely moving solitons. Two distinct, diverging
length scales appear in the limit where the growth rate of trees, $p$,
vanishes. The width of the solitons, $w$, diverges as a power law,
$1/p$, while the average distance between solitons diverges much
faster as $ d \sim \exp({\pi}^2/12p)$.

\end{abstract}

\vskip2pc]

\narrowtext

The Bak-Chen-Tang (BCT) forest fire model\cite{BCT} is a simplified
model of turbulent phenomena.  Propagating fires dissipate or burn
trees at an average rate determined by the tree growth probability,
$p$, representing power fed into the system. The fires are
self-sustaining, with no spontaneous ignition.

The fires exhibit non-trivial spatio-temporal correlations in the slow
driving limit, where the growth rate of trees, $p$, vanishes
\cite{Johansen}.  In two and three dimensions, a ``scale-dependent''
critical behavior emerges \cite{CB}, which is not adequately
understood at present. The fractal dimension of fires depends on the
length scale of observation, up to the correlation length, $\xi$,
where the fires become space filling.  This correlation length
diverges with the inverse growth rate $1/p$ to a power $2/d$ for
$d=2,3$ \cite{CB}.  The behavior in the one dimensional ($d=1$) BCT
model, where the fires are constrained by dimensionality to have a
much simpler structure, has not previously been investigated.

Here we show that in one dimension, as $p \rightarrow 0$, fires
self-organize into a state characterized by propagating, branching and
annihilating solitons.  These are localized objects, compared to their
scale of separation, moving with constant velocity.  Outside of the
solitons only solid forest exists. As an advancing fire moves into
solid forest, each advancing fire is trailed by a cloud of smaller
forests separated by holes, with possibly a few fires.  Each fire
advancing into solid forest together with its trail of finite forests
and fires constitutes a soliton.  New solitons are created by a
process where forward moving solitons emit opposite moving ones.

  The width of the solitons, $w$, diverges as $1/p$.  Within the
solitons, the finite forests, which are connected segments of trees,
have a distribution of sizes, $P(s) \sim 1/s^2$ up to the cut-off
scale $1/p$ imposed by the width of the solitons.  The other
statistical properties are completely determined by the branching rate
for solitons, which can also be calculated analytically.  This
branching process creates new solitons, and must be balanced by the
collision rate of oppositely moving solitons, which annihilates both
solitons. We find that the density of fires remains finite but
decreases extremely fast, $n \sim \exp(-{\pi}^2/12p)$, in the limit $p
\rightarrow 0$.  Thus, even in one dimension, the fires are
self-sustaining with no spontaneous ignition in the slow driving
limit.  Although the solitons are diverging in width, their width
vanishes compared to their separation, and they are well-defined in
the limit $p \rightarrow 0$.

All of this behavior is entirely different from the behavior of the
Drossel-Schwabl (DS) forest fire model\cite{Drossel}, which
self-organizes into a critical state with a power law distribution of
forest fires.  (The BCT model does not have a power-law distribution
of forest fires.)  In the DS model, fires are not self-sustaining but
are injected at a small rate, $f$. The DS model is critical in the
limit $f/p \rightarrow 0$ where a separation in time scales between
burning of entire forests and the growth of trees occurs.  Since the
forests burn down instantly, and the fires are injected rather than
self-sustaining, this limit gives a completely different physical
picture than the propagating fires in the strict $f=0$ problem,
corresponding to the original BCT model.  Solitons, obviously, cannot
appear in the DS model.

The BCT forest fire model is defined as follows. Each site on a
$d-$~dimensional lattice of linear extent $L$, can be in one of three
states: occupied by a fire, occupied by a tree, or empty.  During a
time step, all sites which contain fire burn down, leaving an empty
site, and ignite neighboring trees on the lattice.  Then all empty
sites are independently occupied with trees with probability $p$.
This two step process is repeated indefinitely.  The model can be
studied with either open or periodic boundary conditions, and the
results discussed below are found in either case.  After a transient
period, the system enters a statistically stationary state with a
complex distribution of fires, and forests, which are connected
clusters of trees.  The fire is self-sustaining and fed by the tree
growth process, for $p > 1/\ln L$ (for $d=1$).  Otherwise the fire
dies out completely.

\twocolumn[\hsize\textwidth\columnwidth\hsize\csname @twocolumnfalse\endcsname 
\begin{figure}
\epsfxsize=4.5 truein
\centerline{ \epsffile{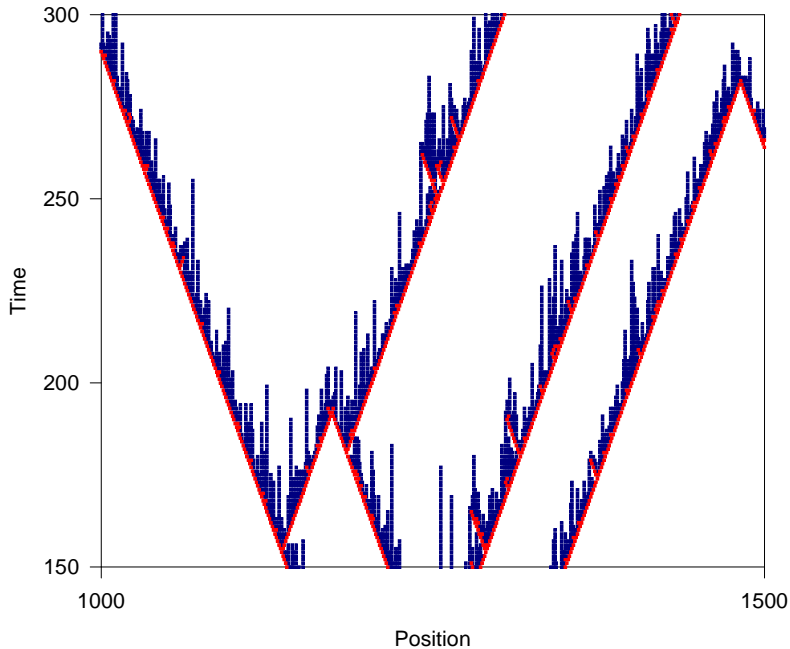} }
\caption{ Solitons in the one-dimensional forest fire model for
$p=0.15$. The red squares are fires, the blue squares are empty
sites. The white sites are trees. This figure shows the essential
dynamical processes that enable an estimate of the contribution of
various terms including collisions between solitons, followed by
complete re-growth of the surrounding forest, and emission of solitons
through back-propagating fires.} 
\end{figure} 
\vskip2pc]

In order to get a clear picture, it is useful to view the space-time
dynamics of the burning process.  The numerical simulations were
initiated with a random distribution of trees and fires.  The
spatio-temporal behavior in the steady-state is shown in Figure 1 for
$p=0.15$. The fires propagate either left or right, moving into solid
forests, with unit velocity. Each such fire leaves a trail of forests,
empty sites, and possibly some fires in their wake.  Each moving
trail, led by a fire advancing with unit velocity into a fully
connected forest, constitutes a soliton.  After some time, of order
$1/p$, the small forests trailing in the wake of the advancing fire
have grown to form a completely connected forest again, with no trace
of the passing fire.  Thus the width of the soliton scales as
$1/p$. For small $p$, such as that shown in Figure 1, the solitons are
well separated from each other and collisions are rare.

When fires moving in opposite directions collide, they annihilate each
other, leaving clusters of small forests. After a time interval of
order $1/p$, the small forests have all healed and joined the large
surrounding forest, and the remnants of both solitons have completely
disappeared.

Now and then, solitons emit solitons propagating in the opposite
direction. This is the fundamental process for the self-organizing
dynamics.  Consider a state with a density $n$ of random placed
solitons, each with a randomly chosen direction. In the stationary
state, the rate of soliton death due to collisions must balance the
rate of new solitons emitted as back-fires. The expected life-time of
a soliton is equal to the average distance between solitons,
$d=1/n$. Assume that each soliton emits back-propagating solitons at a
rate $r$. Stationarity requires $r = 1/d=n$.

In order for a backward-moving fire to survive, and emerge as a
soliton, the new fire must survive a number of time steps, by using
the new, growing forests inside the ``parent'' soliton as stepping
stones, before it can escape from its parent soliton and propagate in
a fully-connected forest.  Since the system is one-dimensional, one
might suspect that this probability will vanish, for any $p<1$, since
sooner or later the newly created fire would always meet an empty site
inside its parent soliton. Actually, for any finite $p$ this is not
the case.

At the first step, where the initial fire moves forward one unit into
a fully connected forest, there is a probability $p$ that this fire
will branch backwards.  In this case, there will be two fires at the
same time at neighboring sites.  This means that a site that was on
fire in the previous time step, $t-1$, is regrown with a tree and set
again on fire at time $t$.  Thus $p$ is the basic branching
probability for new fires to be created.

We will consider the process where the newly emitted fire only moves
backward (see Fig.~1), opposite from the original fire, with no
forward motion, or wandering.  This is the dominant process for $p
\rightarrow 0$. What is the probability that such a newly created fire
will survive one additional time step?  It is $(1-(1-p)^3)$, since it
is now three time steps since the forward-moving, parent fire passed
the advancing position of the new, daughter fire. At the $m$'th time
step for the new fire, the conditional probability of surviving one
more time step is $1-(1-p)^{2m+1}$. Thus, the probability for
surviving $m$ steps is \begin{equation} P_{surv}(m)= p
\prod_{m'=1}^m(1 - (1-p)^{2m'+1}) \quad .
\end{equation}
Taking the limit $m \rightarrow \infty$ and $p \rightarrow 0$ gives
\begin{equation} \lim_{p\rightarrow 0}\ln P_{surv} = {1\over
2p}\int_0^{\infty}\ln(1-e^{-u})du = {-\pi^2 \over 12 p} \quad .
\end{equation}
This is the rate of emitting back-propagating solitons.  From the
previous argument $P_{surv}=n$.  This result has been compared with
numerical simulations as shown in Fig. 2.  Numerical simulation
results for $0.1 < p < 0.3$, and soliton distances ranging from $20$
to $10000$, yield an excellent fit to the form $n=\exp(-C/p)$, with $C
= 0.834$, within 2\% of the exact value ${\pi^2/12}$.

Note that there are two length scales, each diverging with $1/p$, but
in very different ways. One is the width of solitons, $w \sim
1/p$. The other is the much larger distance between solitons,
$d=\exp({\pi}^2/12p)$. Because of the exponentially growing distance
between solitons, it is numerically feasible to simulate the model
only for $p > 0.1$. However, the length scales are already extremely
well separated for these values, establishing the soliton picture.

\begin{figure}
\narrowtext \epsfxsize=2.8truein \centerline{ \epsffile{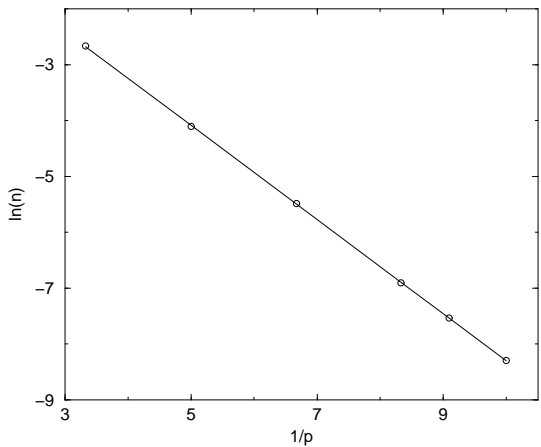} }
\caption{A fit of numerical measurements of the density of fires, $n$,
to the form $n=\exp(-C/p)$, giving $C = 0.834$.}
\end{figure}

Also for any finite $p$ there is always a finite probability that a
soliton will annihilate itself by encountering a rare hole in the
nominally fully connected forest.  However the rate of this
annihilation process for a soliton is $e^{-apd}$, where $a$ is a
constant of order unity.  This is derived by considering that such a
rare hole left over was created by a soliton approximately $d$ time
steps in the past.  This rate vanishes, in the limit $p\rightarrow 0$,
compared to the rate $1/d$ for the pair-wise annihilation described
above.

The distribution of the forest sizes {\it within} the comparatively
narrow solitons has a structure determined entirely by a cascade
process in which isolated clusters of trees form bigger and bigger
forests \cite{Drossel2,PB}.  This is caused by tree growth filling in
the empty sites between neighboring forests. Forests of size $s$ are
both created and destroyed by this same process.  We obtain
analytically a power-law distribution of forests within solitons,
$P(s) \sim 1/s^2$, up to a cut-off, which is the size of the soliton
(of order $1/p$).  The same cascade process occurs in the
Nagel-Schreckenberg traffic model \cite{NS}, where it leads to $1/f$
noise \cite{traffic-jam} and also in the Drossel-Schwabl forest fire
model \cite{PB}, which has spontaneous ignition of fires and
instantaneous burning of forests \cite{Drossel} (and {\it no}
solitons), unlike the model discussed here. It is very important to
distinguish these small forests, forming the micro-structure of
the solitons, from the exponentially larger forests separating
solitons.  In the limit $p \rightarrow 0$, the solitons are point
like, viewed at the large scale of their separation, and fractal-like,
viewed at scales less than their width.

In one dimension, a soliton can propagate only if the forest density
in front of it is unity, so the solitons do not feel the effects of
previous solitons, and, except for branching, there is no interaction
other than at contact.  This behavior is somewhat reminiscent of
one-dimensional Burgers turbulence, where the dynamics is also
governed by colliding solitons, or shocks \cite{turb}.  Persistent,
localized propagating structures that interact via collisions have
also been observed in some cellular automata, which unlike the model
discussed here, are deterministic \cite{crutchfield}.  One distinct
feature of the BCT forest fire model is that a single soliton can emit
other solitons, unlike shocks in Burgers turbulence, or solitons in
cellular automata.

In higher dimensions, the propagating fires interact with each other
at a distance.  This interaction
is  mediated by their effect on the forest density and
tree-tree correlations. The resulting spatio-temporal structure of
dissipation is much more complicated. Thus it is not clear if the
soliton description will remain valid or useful for the higher
dimensional BCT models, or other systems where the interactions
between dissipating structures are long-range.  Nevertheless, the
simple results found here for the one-dimensional BCT forest fire
model suggests that solitons, or shocks, may govern the dynamics of
other self-organized critical models of turbulent, intermittent
phenomena.  In fact, we have observed similar behavior in an interface
depinning model \cite{CPB}.  In this case the sites on fire correspond
to the sites which are moving, and the limit $p \rightarrow 0$
corresponds to the limit where the driving force approaches the
depinning threshold from the moving phase.

PB and MP thank the National University of Singapore for great
hospitality, and working environment.


\end{document}